\documentclass[10pt,twocolumn,letterpaper]{article}

\usepackage{cvpr}              

\usepackage{tipa}
\usepackage{hyperref}

\definecolor{cvprblue}{rgb}{0.21,0.49,0.74}
\usepackage{multirow}

\title{Wav2Sem: Plug-and-Play Audio Semantic Decoupling for 3D Speech-Driven Facial Animation}

\author{Hao Li$^{1,2}$, Ju Dai$^{2,}$\thanks{Corresponding authors}, Xin Zhao$^{1,2}$, Feng Zhou$^{3}$, Junjun Pan$^{1,2,*}$, Lei Li$^{4,5}$\\
$^{1}$Beihang University,  $^{2}$Peng Cheng Laboratory, $^{3}$North China University of Technology, \\$^{4}$University of Washington, $^{5}$University of Copenhagen\\
{\tt\small \{lih09, daij, zhaox03\}@pcl.ac.cn, zhoufeng@ncut.edu.cn, pan\_junjun@buaa.edu.cn, lilei@di.ku.dk } 
}

\begin{document}
\maketitle

\begin{abstract}
In 3D speech-driven facial animation generation, existing methods commonly employ pre-trained self-supervised audio models as encoders. However, due to the prevalence of phonetically similar syllables with distinct lip shapes in language, these near-homophone syllables tend to exhibit significant coupling in self-supervised audio feature spaces, leading to the averaging effect in subsequent lip motion generation. To address this issue, this paper proposes a plug-and-play semantic decorrelation module—Wav2Sem. This module extracts semantic features corresponding to the entire audio sequence, leveraging the added semantic information to decorrelate audio encodings within the feature space, thereby achieving more expressive audio features. Extensive experiments across multiple Speech-driven models indicate that the Wav2Sem module effectively decouples audio features, significantly alleviating the averaging effect of phonetically similar syllables in lip shape generation, thereby enhancing the precision and naturalness of facial animations. Our source code is available at \href{https://github.com/wslh852/Wav2Sem.git}{https://github.com/wslh852/Wav2Sem.git}.
\end{abstract}    
\section{Introduction}
\label{sec:intro}

Speech drive 3D facial animation is widely used in virtual reality, film production, and gaming and has been an active research topic for decades~\cite{ThiesETTN20,VougioukasPP19,WangL0022}. The high correlation between audio and lip shapes makes lip-syncing the most critical metric for evaluating facial animation. Whether utilizing phoneme-based~\cite{XuFMS13} matching techniques or deep learning methods~\cite{FaceDiffuser,VOCA}, the primary focus is on establishing mapping rules between audio signals and facial (particularly lip) movements to achieve accurate lip-syncing.

\begin{figure}[htb]
 \centering
 \includegraphics[width=0.98\columnwidth]{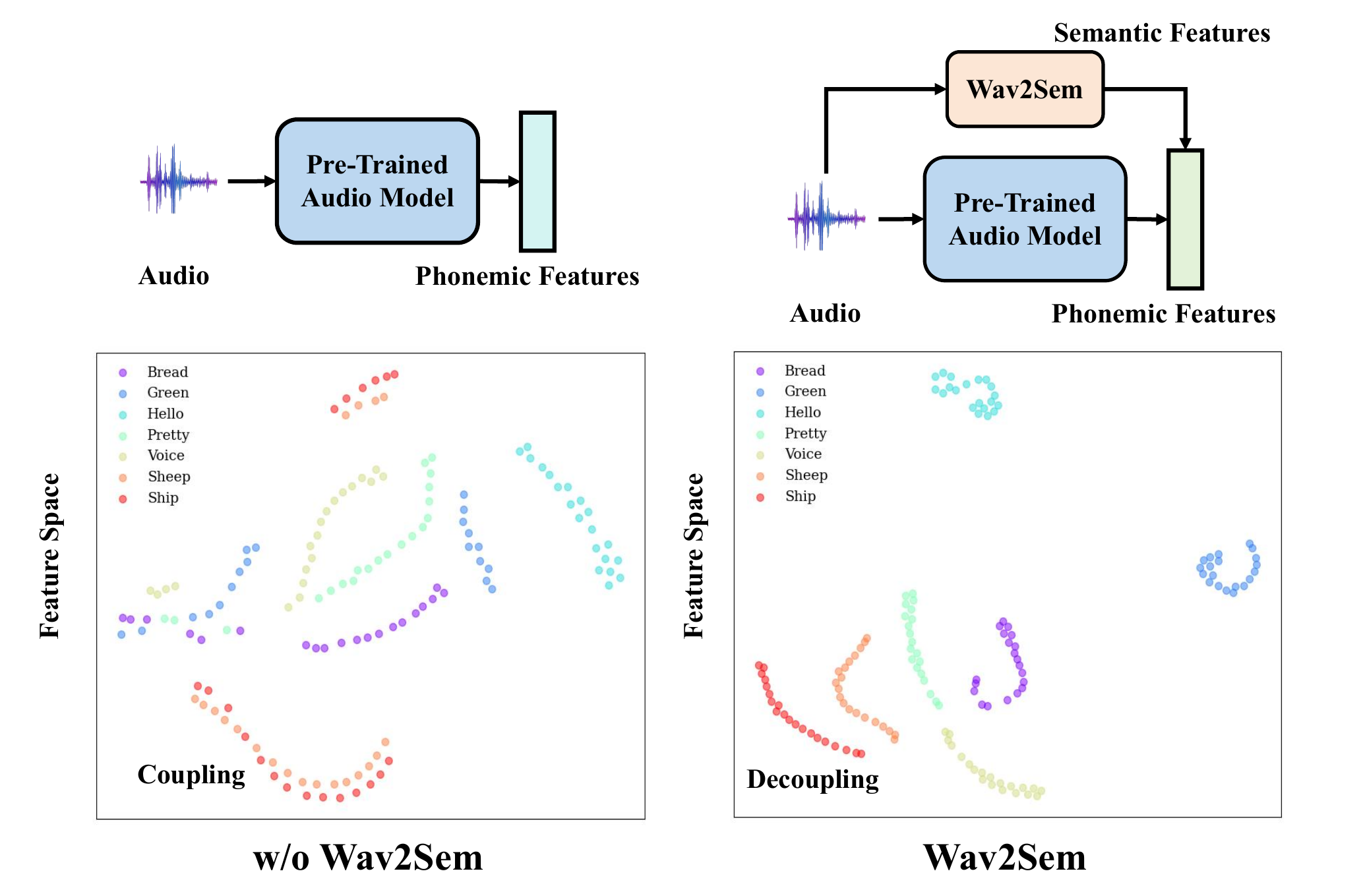}
 \vspace{-4mm}
 \caption{Our plug-and-play Wav2Sem module effectively alleviates the coupling of near-homophonic syllables in pre-trained self-supervised models by extracting semantic features from audio.}
 \label{introl}
\end{figure}

Existing work~\cite{FaceFormer,MeshTalk,FaceDiffuser,thambiraja2023imitator,song2024talkingstyle,yang2025you,sun2023diffposetalk,aneja2024facetalk} typically employs pre-trained self-supervised models for audio encoding, benefiting from the exceptional feature generalization abilities provided by self-supervised learning, which significantly enhances model training efficiency. However, due to the diversity of phoneme combinations in natural language, a considerable number of near-homophonic syllables exist. These syllables are acoustically similar, but their corresponding lip shapes during pronunciation exhibit distinct differences.
For example, When “sheep” produces the long vowel sound /i:/, the mouth opens slightly with a gentle lip parting; whereas with “ship” producing the short vowel sound /\textipa{I}/, the lips remain neutral, and the mouth opens slightly wider. In Figure \ref{introl}, We observe that near-homophonic syllables show significant coupling in the feature space of existing self-supervised models, which leads to the averaging of near-homophonic syllables in the subsequent lip shape generation process, resulting in significant lip-syncing error. Therefore, effectively preventing the coupling of near-homophonic syllables in the feature space becomes a significant challenge.

The prevailing self-supervised audio encoding models, such as HuBERT~\cite{hubert} and Wav2Vec 2.0~\cite{wav2vec}, rely on large-scale unlabeled audio data for training, which often restricts feature modeling to the phoneme-level, lacking effective semantic representation. Some endeavors~\cite{EMAGE,audiogpt2} attempt to incorporate additional text-based features to supplement semantic information. However, this approach requires more complex model designs to address feature alignment and introduces extra input requirements. In practice, homophones are not problematic in daily communication, as listeners can infer their meanings solely through auditory information without needing additional text. This is because, besides phoneme-level information, we rely on understanding the semantic context contained in the audio itself.

Inspired by this, we propose a plug-and-play semantic decoupling module--Wav2Sem. Wav2Sem effectively captures the global semantic features within audio sequences, making it highly adaptable for integration into existing self-supervised audio models to supplement semantic features. Introducing global semantic features extracted from audio enhances the model's deeper understanding of context. Even with simple fusion techniques, it enables effective decoupling of near-homophonic syllables within the feature space. Specifically, Wav2Sem is pretrained on a large-scale text-audio database to align semantic features of text and audio at the sentence level. For the input audio sequences, Wav2Sem employs a multi-layer Temporal Convolutional Network (TCN) to capture local features from the audio signals. It subsequently utilizes a multi-level Transformer~\cite{transformer} to dynamically capture the long-range dependencies among these local features and to learn the mapping from audio to the semantic space of the BERT's~\cite{bert} text. Extensive 
experiments indicate that the semantic features extracted by Wav2Sem can be seamlessly integrated with existing pre-trained self-supervised models, resulting in more expressive feature representations and effectively mitigating the averaging issue presented in lip shape generation tasks. The capabilities of Wav2Sem are also validated for phoneme recognition. Our contributions are as follows:
\begin{itemize}
    \item In speech-driven facial animation, we observe that current pre-trained self-supervised audio models focus on learning phoneme-level features, resulting in significant coupling during the encoding of near-homophonic syllables.
    \item We propose the plug-and-play Wav2Sem, which can directly extract semantic information from audio sequences and seamlessly be integrated into existing pre-trained self-supervised models to decouple near-homophonic syllable features in the feature space.
    \item Extensive experimental validation demonstrates the effectiveness of the proposed Wav2Sem in enhancing facial animation and phoneme recognition performance.
\end{itemize}

\section{Related Works}
\subsection{Speech-driven 3D Facial Animation}
In recent years, speech-driven facial animation has received considerable attention. Traditional approaches~\cite{TaylorKYMKRHM17,cao2005expressive} focus on establishing mapping relationships between phonemes and facial movements, which often rely on predefined facial movements libraries and complex mapping rules. As a result, the generated outcomes frequently lack natural transitions and exhibit rigid characteristics. With the advancements in deep learning, numerous studies~\cite{FaceFormer,CodeTalker,FaceDiffuser,li2023mask,haque2023facexhubert,EMOTE,peng2023emotalk,thambiraja20233diface} have been dedicated to learning the mapping relationships between speech and facial movements from data. For instance, VOCA~\cite{VOCA} is a straightforward and versatile framework that not only enables the generation of facial animations based on speech but also supports cross-identity speech driving. Faceformer~\cite{FaceFormer} is a speech-driven 3D facial animation architecture based on an autoregressive transformer, which utilizes a pre-trained self-supervised audio model to solve the problem of data scarcity in existing audiovisual datasets. Codetalker~\cite{CodeTalker} demonstrates the advantages of transforming speech-driven facial animation into a code-query task in discrete space, which significantly improves the quality of facial motion synthesis. With the development of the diffusion model, FaceDiffuser~\cite{FaceDiffuser} integrates the diffusion mechanism into speech-driven facial generation and achieves more accurate lip synchronization and facial expression. Similarly, LG-LDM~\cite{LG-LDM} employs latent diffusion modeling to ensure that subtle emotional outputs in facial animations are accurately rendered. Mimic~\cite{fu2024mimic} introduces an innovative speech style disentanglement method, which can realize the speech style coding of any subject, so as to synthesize facial animation more realistically. However, these works focus on more complex models to improve the quality of facial generation, ignoring the impact of generating features for pre-trained self-supervised audio models. In particular, languages have many near-homophonic syllables with similar pronunciations but different lip shapes, the incorrect expression of audio feature space affects the naturalness and consistency of the generated results.

\subsection{Audio Encoder in Speech-driver}
The audio encoder aims to convert raw audio signals into embeddings that are easier for the model to process, playing a crucial role in speech-driven facial animation. Earlier studies~\cite{wav2lip,DBLP:conf/bmvc/ChenLLYW21} utilize Mel Frequency Cepstrum Coefficients (MFCC) as the feature representation of audio. However, extracting MFCC from raw audio usually leads to losing high-frequency information and poor robustness. With the development of deep learning, DeepSpeech~\cite{DeepSpeech} is developed, which is an audio feature extractor based on a combination of convolutional and recurrent neural networks, aiming to capture the temporal and frequency domain information in audio more efficiently. However, DeepSpeech is highly dependent on labeled data. In order to reduce the dependence on labeled data, self-supervised learning approaches have received wide attention. Wav2Vec 2.0~\cite{wav2vec} provides an efficient self-supervised learning framework that enables models to automatically learn high-quality audio feature representations by large-scale unlabeled audio data. This approach addresses data dependency and generalization challenges in speech recognition tasks. HuBERT~\cite{hubert} introduces unsupervised clustering and multi-stage training, effectively enhancing feature extraction and task performance. At the same time, it simplifies the complex negative sampling mechanism, providing a more efficient and refined approach to speech representation learning. Due to the scarcity of audiovisual datasets, speech-driven research often uses pre-trained self-supervised models as audio encoders. However, these self-supervised models focus on phonemes and low-level features, lacking overall semantic modeling, which results in the model's inability to distinguish homophones based on context.

\subsection{Multimodal Facial Animation Generation}
With the advancement of multimodal technology, substantial studies~\cite{ao2023gesturediffuclip,liang2024omg,zhang2022motiondiffuse,mughal2024convofusion,zhao2024media2face,DBLP:conf/cvpr/ChhatreDABPBB24} are integrating multimodal information to enhance generation quality. Benjamin et al.~\cite{ElizaldeZR19} propose a multimodal search framework based on co-embeddings of text and audio, aiming to obtain more robust feature representations through a shared embedding space to enhance retrieval performance. Yu et al.~\cite{Yu0L19} generate landmark points intermediate representations by introducing additional textual information, thus improving the final generated facial effects. EMAGE~\cite{EMAGE} is a text and speech-driven character motion generation framework that enables the continuous and natural generation of facial and body movements. These works utilize additional text features to supplement the missing semantic information in the audio. 
It is worth noting that text and audio are two different ways to express semantics, they differ in how the information is conveyed but ultimately convey the same core semantic information. Therefore, it is feasible to directly obtain the corresponding semantic information from the audio. The Wav2Sem module can directly learn corresponding semantic information from the entire audio without the need for additional text assistance. It can be easily integrated into existing facial animation pipelines, significantly improving the model's ability to capture semantic information.
\section{Method}
Existing self-supervised audio modeling mainly focuses on phoneme-level feature modeling. This results in coupling relationships between near-homophonic syllables in the feature space. Regarding paired text and audio, they are just different expressions, while the intrinsic semantics are the same. Therefore, this work aims to learn global semantic features from audio signals and achieve feature decoupling from existing audio self-supervised models by introducing global semantic features for a more robust feature representation. 
To achieve this goal, We propose the plug-and-play Wav2Sem, which can effectively extract semantic features corresponding to audio signals. By integrating corresponding semantic features into existing audio self-supervised models, the coupling of near-homophonic syllables in the feature space can be effectively decoupled, thereby improving the representation of audio features.

\begin{figure*}[htb]
 \centering 
 \includegraphics[width=1.85\columnwidth]{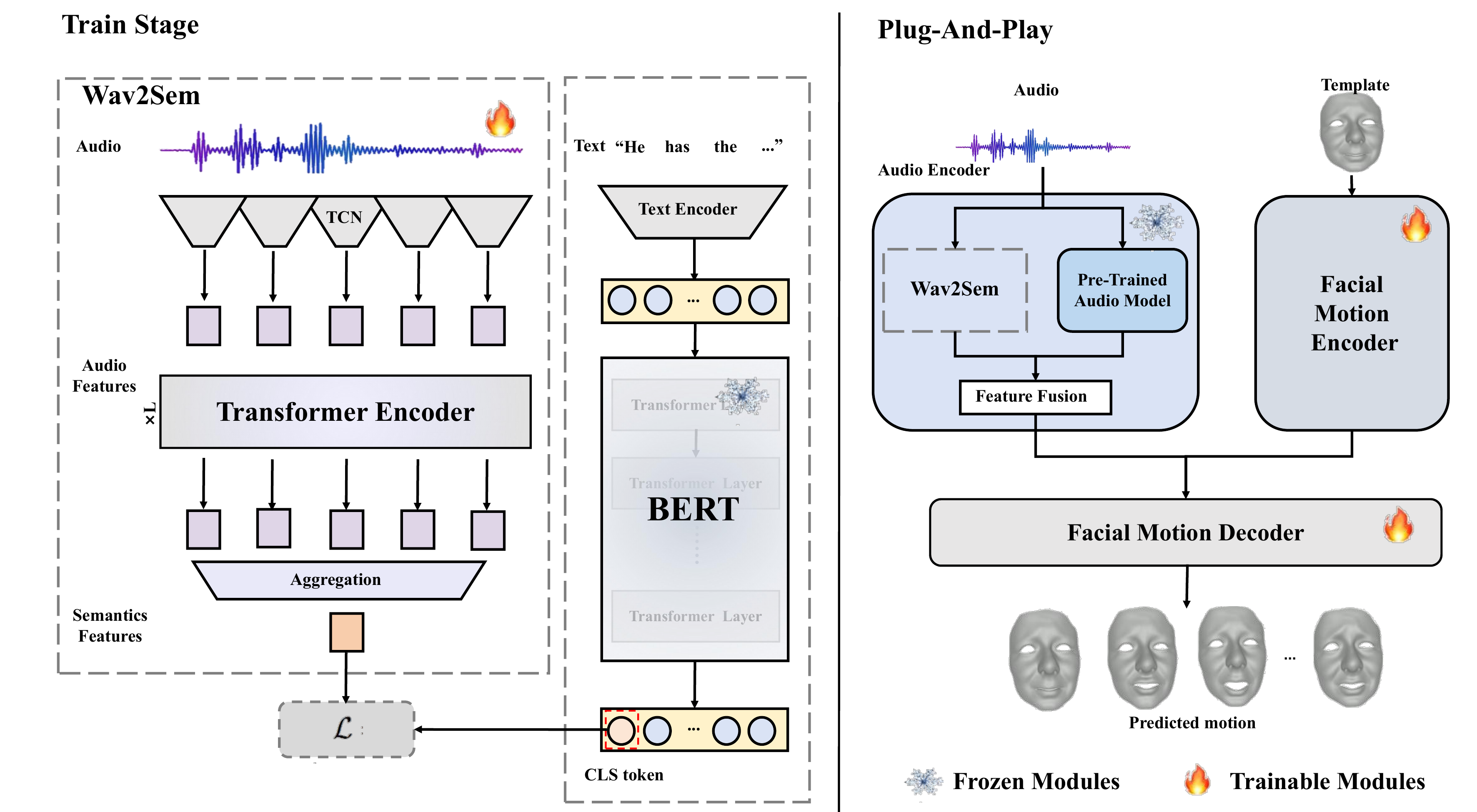}
 \vspace{-3mm}
 \caption{Overview of Wav2Sem, which is trained on a large dataset of text-audio pairs. Given input audio signals, Wav2Sem extracts semantic information using a Temporal Convolutional Network (TCN) and a Transformer-based encoder. The learned semantic information is aligned with the text semantics in the BERT space. Integrating Wav2Sem into the audio encoder of an arbitrary speech-driven facial animation framework enables the decoupling of the near-homophonic syllable feature space, resulting in varying performance enhancement.}
 \label{pipline}
\end{figure*}

\subsection{Semantic Space}
A defined and complete semantic space is crucial for the performance of the model. In fact, building a complete semantic space from audio is difficult because audio signals contain rich acoustic features. For the same content, the audio signal can be expressed by different acoustic features, and this diversity complicates the construction of a consistent speech feature space in audio. Compared to continuous audio, discrete and structured text has clear semantics boundaries, and it is easy to represent the semantics of each word by embedding vectors. However, aligning audio features with word-level semantic features is challenging, and word-level semantic features usually fail to comprehensively capture the overall semantics of a sentence and lack contextual associations. The proposed Wav2Sem aims to achieve high-dimensional alignment of audio and text, so we use the semantic representation of sentences in BERT~\cite{bert} as our semantic space, and it can be formed as:
\begin{equation}
\setlength\abovedisplayskip{3pt}
\setlength\belowdisplayskip{3pt}
   \mathbf{F}_{CLS} = \mathbf{BERT}(x_1,x_2,…,x_M),
\end{equation}
where $\mathbf{F}_{CLS}$ denotes the sentence embedding vector obtained after encoding the pre-trained BERT model. $x_1,x_2,…,x_M$ are the individual words in the input sentence, and $M$ is the total number of words.

\subsection{Wav2Sem Learning}
Wav2Sem consists of a Temporal Convolutional Network (TCN) and a multilayer transformer encoder. The former aims to capture local information of the input audio, and the latter is responsible for integrating the local information to obtain globally aware semantic representations of the input audio.
For a given audio $\mathbf{A} = (a_1, ... , a_K)$, it is first processed by the TCN component:
\begin{equation}
\setlength\abovedisplayskip{3pt}
\setlength\belowdisplayskip{3pt}
   \mathbf{Z} = \operatorname{TCN}(\mathbf{A}), 
\end{equation}
where $\mathbf{Z} \in \mathbb{R}^{N \times C}$ is the local feature information, $N$ is the output length, and $C$ is the feature dimension.

Then, we learn the contextual relationships through the multilayer transformer encoder. The process is as follows:
\begin{equation}
\setlength\abovedisplayskip{3pt}
\setlength\belowdisplayskip{3pt}
\mathbf{\hat{Z}}^{l} = \mathbf{{Z}}^{l-1} +\operatorname{MHSA}(\operatorname{LN}(\mathbf{Z}^{l-1})),
\end{equation}
\begin{equation}
\setlength\abovedisplayskip{3pt}
\setlength\belowdisplayskip{3pt}
\mathbf{Z}^{l} = \mathbf{\hat{Z}}^{l}+\operatorname{MLP}(\operatorname{LN}(\mathbf{\hat{Z}}^{l})),
\end{equation}
where MHSA and LN refer to the multi-head self-attention and layer normalization in the transformer. $\mathbf{Z}^{l-1}$ represents the output from ($l-1$)-th layer, $\mathbf{\hat{Z}}^{l}$ is the intermediate result.

Finally, Wav2Sem aggregates contextual information as semantic information output for the whole audio sequence:
\begin{equation}
\setlength\abovedisplayskip{3pt}
\setlength\belowdisplayskip{3pt}
\mathbf{F}_{s} = \frac{1}{N}\sum\limits_{i=0}\limits^N \mathbf{Z}_i,
\end{equation}

\subsection{Semantic Fusion}
Our goal is to build a plug-and-play audio semantic decoupling module. In order to reduce the complexity of integration and enable users to easily apply the module to diverse self-supervised audio encoders, our fusion module adopts a simple and intuitive design. For the input audio $\mathbf{A}$, the phoneme feature $\mathbf{F}_p$ encoded by the pre-trained self-supervised audio encoder $\mathbb{M}_A$ can be expressed as:
\begin{equation}
\setlength\abovedisplayskip{3pt}
\setlength\belowdisplayskip{3pt}
\mathbf{F}_p= \mathbb{M}_A(\mathbf{A})
\end{equation}
where $\mathbf{F}_p \in \mathbb{R}^{N' \times C}$, $N'$ is the output length, and $C$ is the embedding dimension.

We achieve semantic fusion by integrating the audio semantic feature into the phoneme feature through two linear transformations and the summing operation:
\begin{equation}
\setlength\abovedisplayskip{3pt}
\setlength\belowdisplayskip{3pt}
\mathbf{F}_d = \operatorname{FC}(\operatorname{FC}(\mathbf{F}_{s})+\mathbf{F}_p)
\end{equation}
 where $\mathbf{F}_{s} \in \mathbb{R}^{1 \times C}$ is the audio semantic feature, $\mathbf{F}_d \in \mathbb{R}^{N' \times C}$ is the decoupled audio feature.

\subsection{Optimizing Function}
Wav2Sem focuses on aligning audio semantics with text semantics at the sentence level, which utilizes the understanding of the overall audio to decouple near-homophonic syllables in the feature space. We optimize Wav2Sem on the large-scale audio transcription dataset LibriSpeech-960\cite{PanayotovCPK15}. The L1 loss function is applied to calculate the difference between audio and text semantics features:
\begin{equation}
\setlength\abovedisplayskip{3pt}
\setlength\belowdisplayskip{3pt}
\mathcal{L} = ||\mathbf{F}_{CLS} - \mathbf{F}_{s}||_1
\end{equation}
where $\mathbf{F}_{CLS}$ is the semantic feature of the sentence text, $\mathbf{F}_{s}$ is the semantic feature of the audio.


\subsection{Training Details}
 We provide two versions of Wav2Sem based on the BERT CLS token ($\text{BERT}_c$), namely $\text{Wav2Sem}_c$, and the average of tokens generated by BERT ($\text{BERT}_m$), namely $\text{Wav2Sem}_m$. The two versions have the same structure and contain 7-layer TCN blocks and 12-layer transformer blocks. Each TCN block has 512 channels, with an output frequency of 49 Hz and a stride of about 20 ms between each sample. Each transformer block contains 8 attention heads, and the internal dimension of MLP is 3072. Wav2Sem is developed based on PyTorch and trained on RTX A6000 Ada 
 with 200 epochs. The optimizer is chosen to be Adam~\cite{KingmaB14}, the learning rate is set to $1 \times 10^{-4}$, and the batch size is 1. 

After finishing the optimization of Wav2Sem, the model parameters are frozen. Then we integrate Wav2Sem into the pre-trained self-supervised audio encoders of different baseline models. The hyperparameters provided in the original papers are applied to optimize all networks.

\section{Experiments}

\subsection{Datasets}
Wav2Sem is optimized on the large-scale speech transcription dataset Librispeech~\cite{PanayotovCPK15}. For speech-driven facial animation tasks, we train and test on two publicly available 3D facial datasets: BIWI~\cite{FaceFormer} and VOCASET~\cite{VOCA}. Additionally, To verify the transfer ability of Wav2Sem, we validate the improvement brought by the Wav2Sem on the phoneme recognition task using the TIMIT dataset~\cite{TIMIT}.

\textbf{Librispeech}~\cite{PanayotovCPK15} is a large-scale speech recognition dataset, which includes audio with various styles and accents, along with corresponding text sampled at a frequency of 16 kHz. Additionally, Librispeech provides several subsets with different amounts of annotations (100 hours, 360 hours, 500 hours) to meet the needs of various experiments. In summary, it consists of 960 hours of training data, 11 hours of validation data, and 11 hours of testing data. The division of these subsets allows researchers to train and evaluate the task with varying scales and complexities.

\textbf{BIWI}~\cite{FaceFormer} is a corpus of emotional speech and dense, dynamic 3D face geometry, consisting of 14 subjects who are asked to read 40 English sentences each in neutral and emotional contexts. The dataset captures 3D face geometry at 25 FPS (frames per second) and contains 23,370 vertices with an average duration of 4.67 seconds. In experiments, BIWI is divided into a training set (192 sentences), a validation set (24 sentences), and two testing sets: BIWI-Test-A (24 sentences) and BIWI-Test-B (32 sentences).

\textbf{VOCASET}~\cite{VOCA} contains audio and 4D scan pairs captured from 6 female and 6 male subjects. Each subject records 40 English sentences of 3 to 5 seconds from various standard protocols to maximize speech diversity. Among the dataset, 5 sentences are shared across all subjects, 15 sentences are spoken by 3 to 5 subjects (totaling 50 unique sentences), and the remaining 20 sentences are spoken by only 1 to 2 subjects (totaling 200 unique sentences). In experiments, we use the same training, validation, and testing segmentation as in VOCA~\cite{VOCA}.

\textbf{TIMIT}~\cite{TIMIT} is an acoustic-phonetic continuous speech corpus sampled at 16 kHz and containing 6300 sentences. These sentences are read aloud by 630 speakers from the eight major dialect regions of the United States, and each reads 10 sentences. All sentences are manually segmented and labeled at the phoneme level, aiming to provide rich acoustic data for speech recognition and related research.

\subsection{Baseline Models}
We select six representative baseline models to validate the effectiveness of Wav2Sem, which include CNN-based VOCA~\cite{VOCA}, Transformer-based FaceFormer~\cite{FaceFormer}, VQVAE-based CodeTalker~\cite{CodeTalker}, TCN-based UniTalker~\cite{UniTalker}, diffusion model-based FaceDiffuse~\cite{FaceDiffuser}, and latent diffusion-based LG-LDM~\cite{LG-LDM}. Since these models are designed with different architectures, it enables us to comprehensively evaluate the performance of Wav2Sem in audio and facial animation generation tasks.

\subsection{Quantitative Evaluation}
We use Lip Vertex Error (LVE) to measure the accuracy of lip movement. The Upper-face Dynamics Deviation (FDD) provides a statistical perspective on motion variation. 
Additionally, we provide the Mean Vertex Error (MVE) metric, as we are interested not only in lip synchronization but also in overall facial movement.

\textbf{LVE} computes the error between the ground truth and the generated lip vertices over the whole sequence.
\begin{equation}
\setlength\abovedisplayskip{1pt}
\setlength\belowdisplayskip{1pt}
\mathrm{LVE} = \frac{1}{T} \sum_{t=1}^{T}  || \mathbf{V}^{(t)}_{\mathrm{lip}} - \hat{\mathbf{V}}^{(t)}_{\mathrm{lip}} ||_2,
\end{equation}
where $T$ denotes the number of frames in the sequence, $\mathbf{V}^{(t)}_{\mathrm{lip}}$ is the position of the real lip vertex in the $t$-th frame, $\hat{\mathbf{V}}^{(t)}_{\mathrm{lip}}$ is the generated lip vertex, and $||\cdot||_2$ denotes $L2 $ norm.

\textbf{MVE} evaluates the average error between the ground truth (GT) and generated facial vertex positions:
\begin{equation}
\setlength\abovedisplayskip{1pt}
\setlength\belowdisplayskip{1pt}
\mathrm{MVE} = \frac{1}{T} \sum_{t=1}^{T} ||\mathbf{V}^{(t)}_{\mathrm{face}} - \hat{\mathbf{V}}^{(t)}_{\mathrm{face}}||_2,
\end{equation}
where $\mathbf{V}^{(t)}_{\mathrm{face}}$ is the GT position of all facial vertices at frame $t$. $\hat{\mathbf{V}}^{(t)}_{\mathrm{face}}$ is the corresponding predicted positions.

\textbf{FDD}~\cite{CodeTalker} measures the difference in temporal dynamics between the ground truth and generated motion sequences for vertices in the upper face region. It can be described as:
\begin{equation}
\setlength\abovedisplayskip{1pt}
\setlength\belowdisplayskip{1pt}
\hspace{-3mm}
\mathrm{FDD} = \frac{1}{|S_U|}\sum_{v \in S_U}(\mathrm{dyn}(\mathbf{V}_\mathrm{u-face}^t)- \mathrm{dyn}(\hat{\mathbf{V}}_\mathrm{u\!-\!face}^t)),
\end{equation}
where $\mathbf{V}_\mathrm{u-face}^t$ and $\hat{\mathbf{V}}_\mathrm{u-face}^t$ are the GT and predicted 
upper face vertices at frame $t$, respectively. 
$S_U$ is the index set of upper-face vertices. dyn$(\cdot)$
denotes the standard deviation of the element-wise L2 norm
along the temporal axis.

\begin{table*}[htp]
\centering
\caption{Quantitative evaluation for Wav2Sem by integrating it into different frameworks on VOCASET and BIWI.}
\setlength\tabcolsep{4pt}
\vspace{-2mm}
\resizebox{1.63\columnwidth}{!}{%
\begin{tabular}{c  c  c c c c c c }
\hline
\multirow{3}{*}{Method} & \multirow{3}{*}{Audio Encoder} & \multicolumn{3}{c}{VOCASET} & \multicolumn{3}{c}{BIWI} \\
\cline{3-8}
&   &MVE & LVE & FDD & MVE & LVE & FDD      \\
&   & ($\times 10^{-5}$) & ($\times 10^{-5}$) & ($\times 10^{-7}$) & ($\times 10^{-3}$) & ($\times 10^{-4}$) & ($\times 10^{-5}$)    \\ 
\hline 
VOCA~\cite{VOCA}  &\multirow{3}{*}{DeepSpeech} &6.1571 &4.9245&4.8447 &8.3606&6.7158&7.5320\\
VOCA+Wav2Sem$_m$  &  &6.0358 &4.9032&4.7926 &8.3217&6.6982&7.4963\\
VOCA+Wav2Sem$_c$  &  &\textbf{6.0015} &\textbf{4.8915}&\textbf{4.7814} &\textbf{8.3175}&\textbf{6.6821}&\textbf{7.3571}\\
\hline 
Faceformer~\cite{FaceFormer}  &\multirow{3}{*}{Wav2Vec 2.0} &5.1558 &4.1090&4.6675 &7.1658 &4.9847&5.0972\\
Faceformer+Wav2Sem$_m$ &   &5.1169 &4.0654&4.6125 &7.1447 &4.9724&5.0521\\
Faceformer+Wav2Sem$_c$ & 	 &\textbf{5.0962} &\textbf{3.9891}&\textbf{4.5317} &\textbf{7.0956} &\textbf{4.9571}&\textbf{4.9157}\\
\hline
Codetalker~\cite{CodeTalker}  &\multirow{3}{*}{Wav2Vec 2.0} &4.6132 &3.9445&4.4522 &7.3980 &4.7914&4.1170\\
Codetalker+Wav2Sem$_m$  &  &\textbf{4.4714} &\textbf{3.8714}&\textbf{4.3517} &7.2782 &4.7847&3.9925\\
Codetalker+Wav2Sem$_c$  &	 &4.5762 &3.8838&4.3654 &\textbf{7.2519} &\textbf{4.7751}&\textbf{3.9181}\\
\hline
UniTalker~\cite{UniTalker} &\multirow{3}{*}{Wav2Vec 2.0}  &4.1271 &3.5416 &4.7125 &6.6104 &4.0213&4.1296\\
UniTalker+Wav2Sem$_m$ &    &4.0988 &3.1521 &4.5314 &6.5121 &3.9213 &4.0871\\
UniTalker+Wav2Sem$_c$ &    &\textbf{4.0973} &\textbf{3.1476} &\textbf{4.5174} &\textbf{6.5101} &\textbf{3.9112} &\textbf{4.0214}\\
\hline
FaceDiffuse\cite{FaceDiffuser}  &\multirow{3}{*}{HuBERT} &4.3651 &3.7924 &4.2647 &6.8088 &4.2985&3.9101\\
FaceDiffuse+Wav2Sem$_m$  & &\textbf{4.3347} &\textbf{3.7628} &\textbf{4.2157} &6.7216 &4.2816&3.9004\\
FaceDiffuse+Wav2Sem$_c$  & &4.3493 &3.7739 &4.2523 &\textbf{6.5725}&\textbf{4.2521}&\textbf{3.8635}\\
\hline
LG-LDM~\cite{LG-LDM}  &\multirow{3}{*}{HuBERT} &3.7162 &3.7925&3.9154 &7.7298 &4.9869&4.3637\\
LG-LDM+Wav2Sem$_m$  & &\textbf{3.6927} &\textbf{3.7717}&\textbf{3.8831} &7.7521 &4.9626&4.3437\\
LG-LDM+Wav2Sem$_c$  &	&3.7042 &3.7863&3.9014 &\textbf{7.6952} &\textbf{4.9258}&\textbf{4.3121}\\
\hline
\end{tabular}
 }
\label{quatitative}
\end{table*}

We compare the performance of different models
in Table \ref{quatitative}. In experiments, we insert our plug-and-play Wav2Sem into the audio encoders of these models without changing the overall structure. We validated the effectiveness of two different semantic expression designs for Wav2Sem (see 4.6 for details). Notably, the semantic features extracted from audio by two different Wav2Sem demonstrate excellent adaptability with both the state-of-the-art Wav2Vec 2.0~\cite{wav2vec} and HuBERT~\cite{hubert} models. The integration of Wav2Sem significantly enhances the feature encoding capability of each model. Evaluated by LVE, MVE, and FDD metrics, the results show that all the models significantly improve these metrics, further proving the effectiveness of our module. This suggests that our Wav2Sem not only directly improves the model's facial dynamics generation performance but also achieves a high degree of co-coding of speech and facial features without adding additional complexity.

\begin{figure}[htb]
 \centering 
 \includegraphics[width=0.94\columnwidth]{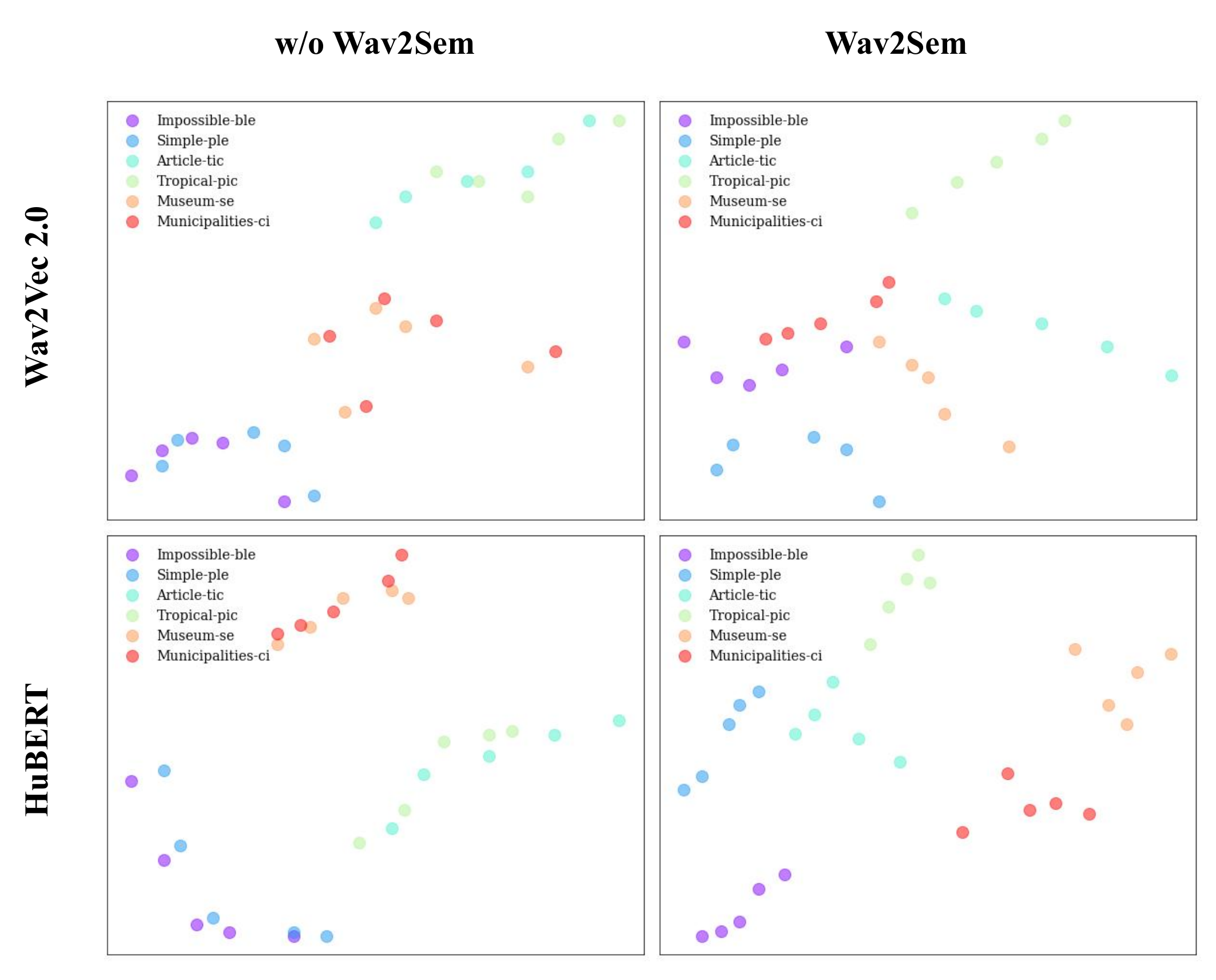}
 \vspace{-2mm}
 \caption{T-SNE comparison for near-homophonic syllables.}
 \label{fdecoupling}
\end{figure}

\subsection{Audio Feature Space Decoupling}
To verify the improvement of Wav2Sem in the near-homophonic syllable feature space, We first extract homophones and near-homophonic syllables from VOCASET text using phoneme similarity rules~\cite{hixon2011phonemic} based on CMU phoneme dictionary (\url{http://www.speech.cs.cmu.edu/cgi-bin/cmudict}). Subsequently, with the help of GPT-4o mini, we obtain 90 pairs of words containing homophones and near-homophonic syllables.

\begin{table}[htp]
\centering
\caption{Comparison of decoupling features in word-level.}
\setlength\tabcolsep{4pt}
\vspace{-2mm}
\begin{tabular}{ccc}
\hline
Model & w/o Wav2Sem & Wav2Sem \\
\hline
Wav2Vec 2.0\cite{wav2vec} &0.0397  &0.0701 \\
\hline
HuBERT~\cite{hubert}     &0.2689  &0.2909 \\
\hline
\end{tabular}
\label{decoup}
\end{table}

As shown in Table \ref{decoup}, we first use the L2 norm to evaluate the decoupling of Wav2Sem at the word level. We map word pairs to the corresponding feature space using Wav2Vec 2.0~\cite{wav2vec} and HuBERT~\cite{hubert}, then calculate the difference between feature vectors. By comparing the word-level differences with and without Wav2Sem, 
We noticed that in the two different feature space representations of Wav2Vec 2.0 and HuBERT, the feature vectors decoupled by Wav2Sem showed significant differences in the feature space, indicating that Wav2Sem has a significant effect on optimizing feature representation.

To provide a more intuitive display of phoneme-level decoupling, We crop the corresponding near-homophonic syllable audio and present the visualization results of the feature vectors of three groups of near-homophonic syllables after T-SNE dimensionality reduction. As shown in Figure \ref{fdecoupling}, we observe that Wav2Vec 2.0 and HuBERT have difficulty distinguishing consonant clusters /pl/ and /bl/ and fast meta consonant combinations /t\textipa{I}k\textschwa l/ /p\textipa{I}k\textschwa l/ in the feature space, while the incorporation of Wav2Sem semantic information significantly improves the discrimination of near-homophonic syllables. This effectively demonstrates the feature decoupling of Wav2Sem at the phoneme level.

\begin{figure*}[htb]
 \centering 
 \includegraphics[width=1.75\columnwidth]{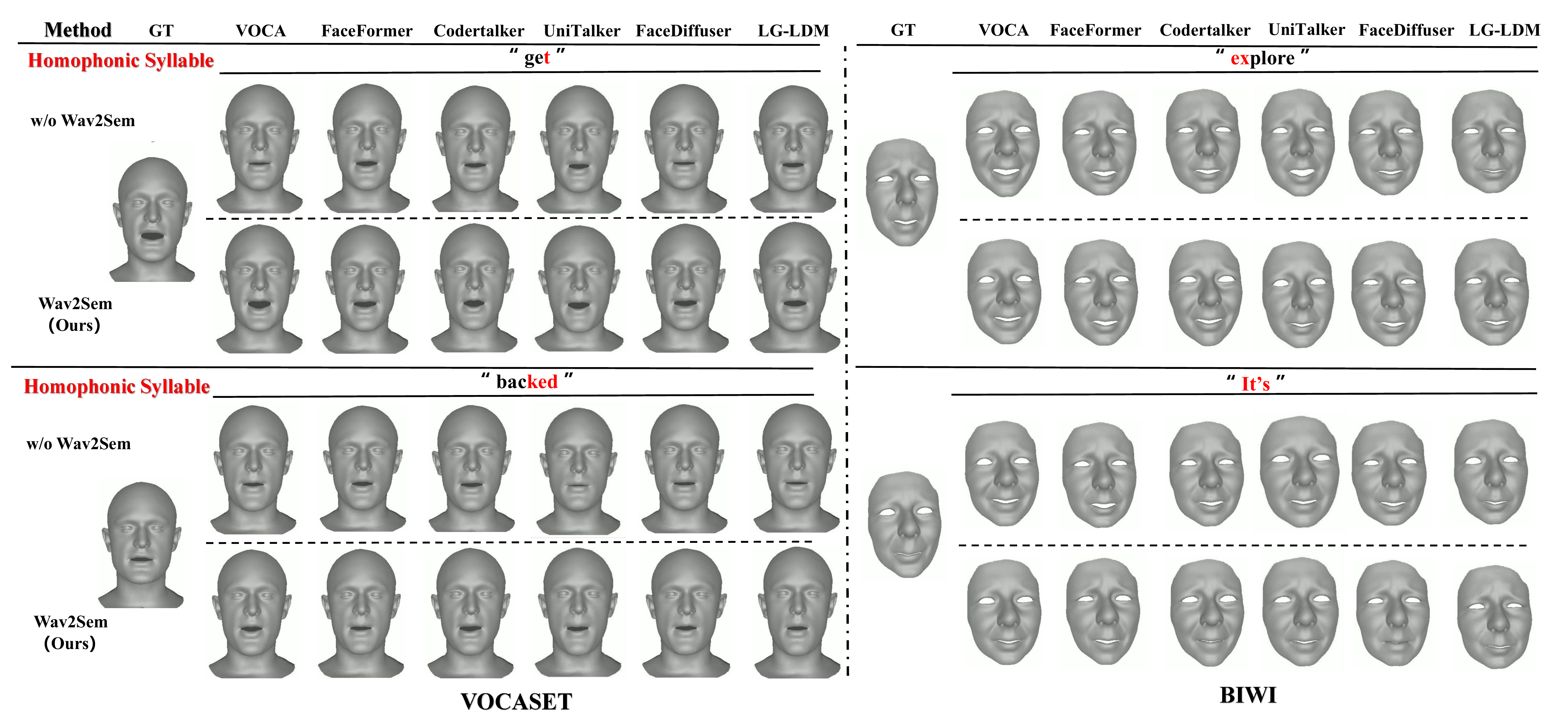}
 \vspace{-3mm}
 \caption{Evaluations of facial motions with and without Wav2Sem 
 for different methods on VOCASET (left) and BIWI (right).}
 \label{Qualitative evaluation}
\end{figure*}

\subsection{Qualitative Evaluation}
For fair comparisons, we assign the same speech style as the conditional input to six baseline models. In Figure \ref{Qualitative evaluation}, we present visualizations of two sets of near-homophonic syllables in VOCASET~\cite{VOCA} and BIWI~\cite{FaceFormer}. For consonants /t/ and /d/, their pronunciations are similar, and there is severe coupling in the feature space of the self-supervised audio model (left side of Figure \ref{fdecoupling}). However, due to the reason for linking, the mouth opening of /t/ in get is more significant than that of /d/ in backed. It can be observed that without semantic decoupling, the facial postures generated by /t/ and /d/ tend to be more averaged (lines 1 and 3). Wav2Sem provides ample semantic information in audio, decoupling near-homophonic syllables in the feature space and generating facial movements with clear discrimination (lines 2 and 4). 
Similarly, we demonstrate the near-homophonic syllables /\textipa{I}ks/ and /\textipa{I}ts/ in the BIWI~\cite{FaceFormer}. Due to the different places of articulation for /k/ and /t/, /\textipa{I}ks/ causes the mouth to open slightly more, while /\textipa{I}ts/ appears more compact. However, due to coarticulation, their phonetic features are coupled in the feature space, which causes the mouth shape for /\textipa{I}ts/ to resemble that of  /\textipa{I}ks/ (lines 1 and 3). It can be observed that the inclusion of Wav2Sem significantly improves this erroneous phenomenon.

\begin{figure}[htb]
 \centering 
 \includegraphics[width=0.9\columnwidth]{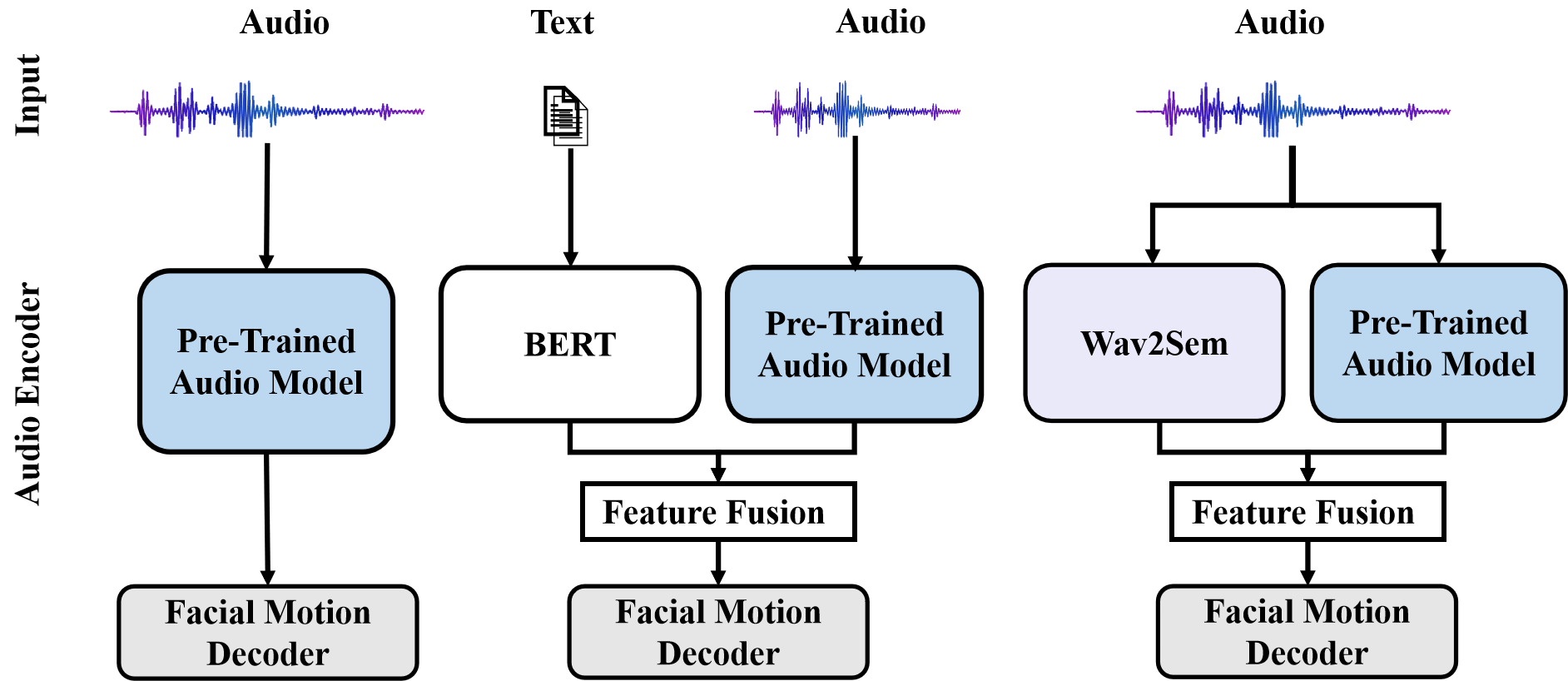}
 \caption{Visualizations of different audio encoding structures}
 \label{fablation}
\vspace{2mm}
 \centering 
 \includegraphics[width=\columnwidth]{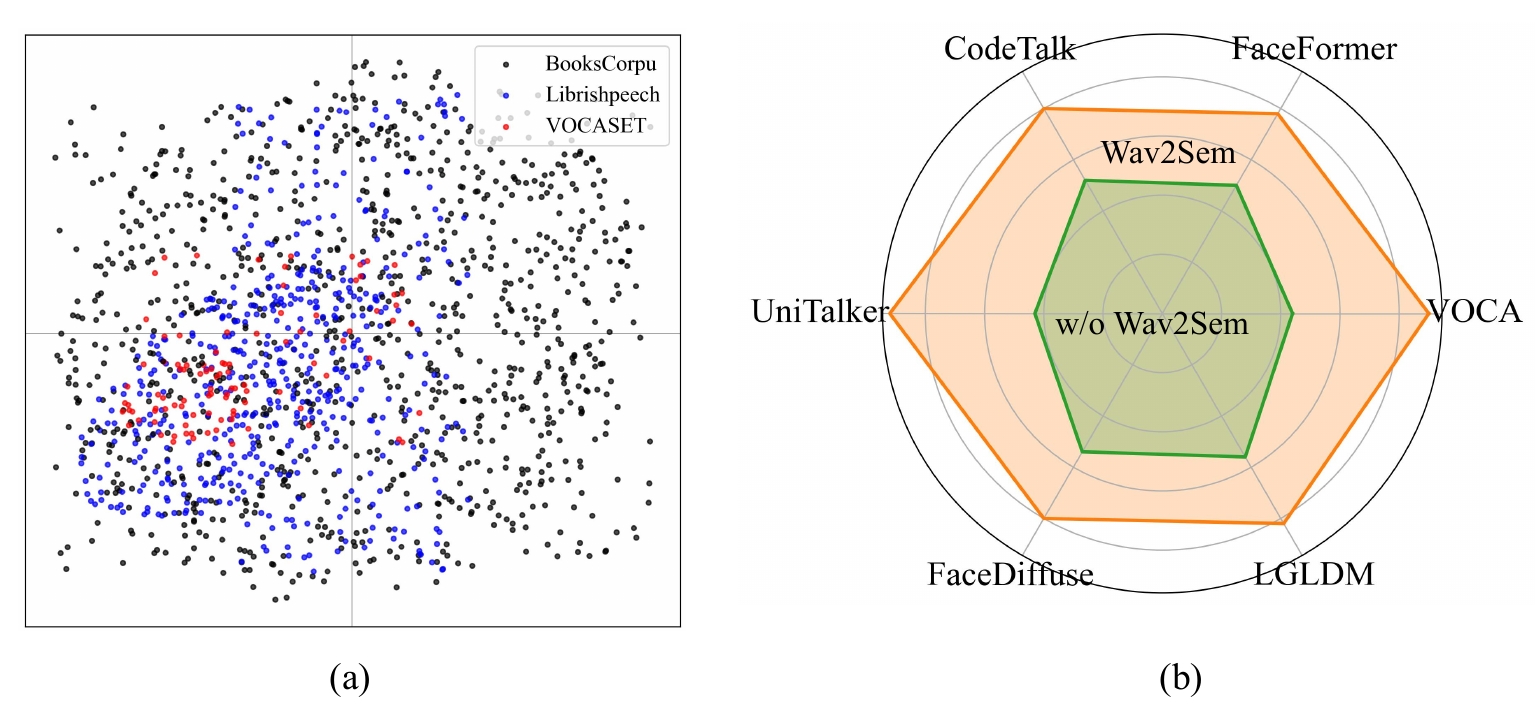}
 \vspace{-6mm}
 \caption{Visualizations of (a) T-SNE distributions of different datasets and (b) user study evaluation on generated facial motions of different methods with and without Wav2Sem.}
 \label{userstudy}
\end{figure}

\subsection{Ablation Studies}

\textbf{Semantic Representation}
There are various ways to represent the semantics in a text, and one of the most common ways is to use the CLS token in BERT for the overall semantic representation. Another effective representation is to average all text tokens, which captures the contribution of individual tokens of the text. During training, we use the pseudo-labels provided by the CLS token and the average of all tokens generated by BERT~\cite{bert} to train two different versions of Wav2Sem, namely $\text{Wav2Sem}_c$ and $\text{Wav2Sem}_m$, and verify which representation is superior. Overall, in the smaller dataset VOCASET~\cite{VOCA}, $\text{Wav2Sem}_m$ shows more advantages than $\text{Wav2Sem}_c$, as illustrated in Table \ref{ablation}. However, in larger BIWI~\cite{FaceFormer}, CLS tokens exhibit outstanding performance, as seen in Table \ref{quatitative}. We believe that CLS tokens can fully capture global information, but due to limited data, the training process is prone to overfitting. In contrast, using the average value of all tokens as the feature representation has better anti-interference performance, so $\text{Wav2Sem}_m$ shows better performance in smaller datasets.

\begin{table}[htp]
\centering
\caption{Ablation experiment on VOCASET, we compare the effects of different semantic decouplings.}
\setlength\tabcolsep{4pt}
\vspace{-2mm}
\resizebox{0.92\columnwidth}{!}{
\begin{tabular}{cccc}
\hline
\multirow{2}{*}{Method} &MVE  &LVE  &FDD  \\
                & ($\times 10^{-5}$) & ($\times 10^{-5}$) & ($\times 10^{-7}$)\\
\hline
VOCA~\cite{VOCA}   &6.1571 &4.9245&4.8447 \\
VOCA+BERT$_m$   &6.0614 &4.9041&4.7927 \\
VOCA+BERT$_c$   &6.0468 &4.8995&4.7842 \\
VOCA+Wav2Sem$_m$   &6.0358 &4.9032 &4.7926 \\
VOCA+Wav2Sem$_c$   &\textbf{6.0015} &\textbf{4.8915} &\textbf{4.7814} \\
\hline
Faceformer~\cite{FaceFormer}   &5.1558 &4.1090&4.6675 \\
Faceformer+BERT$_m$   &5.1027 &4.0872&4.6071 \\
Faceformer+BERT$_c$  &5.1001 &4.0241&4.6241 \\
Faceformer+Wav2Sem$_m$   &5.1169 &4.0654 &4.6425 \\
Faceformer+Wav2Sem$_c$  &\textbf{5.0962} &\textbf{3.9891} &\textbf{4.5317} \\
\hline
Codetalker~\cite{CodeTalker}   &4.6132 &3.9445&4.4522 \\
Codetalker+BERT$_m$  &4.5689 &3.9198&4.4032 \\
Codetalker+BERT$_c$   &4.5739 &3.9217&4.4142\\
Codetalker+Wav2Sem$_m$   &\textbf{4.4714} &\textbf{3.8714} &\textbf{4.3517} \\
Codetalker+Wav2Sem$_c$   &4.5762 &3.8838 &4.3654 \\
\hline
UniTalker~\cite{UniTalker}    &4.1271 &3.5416 &4.7125  \\
UniTalker+BERT$_m$   &4.1086 &3.5261 &4.7253 \\
UniTalker+BERT$_c$   &4.1024 &3.5011 &4.7011 \\
UniTalker+Wav2Sem$_m$   &4.0988 &3.1521 &4.5314 \\
UniTalker+Wav2Sem$_c$   &\textbf{4.0973} &\textbf{3.1476} &\textbf{4.5174} \\
\hline
FaceDiffuse~\cite{FaceDiffuser}   &4.3651 &3.7924&4.2647 \\
FaceDiffuse+BERT$_m$    &4.3514 &3.7861&4.2531\\
FaceDiffuse+BERT$_c$   &4.3816 &3.8066&4.2798 \\
FaceDiffuse+Wav2Sem$_m$   &\textbf{4.3347} &\textbf{3.7628} &\textbf{4.2157} \\
FaceDiffuse+Wav2Sem$_c$   &4.3493 &3.7739 &4.2523 \\
\hline
LG-LDM~\cite{LG-LDM}   &3.7162 &3.7925&3.9154 \\
LG-LDM+BERT$_m$   &3.7557 &3.8614&3.9617 \\
LG-LDM+BERT$_c$   &3.7141 &3.7891&3.9054 \\
LG-LDM+Wav2Sem$_m$   &\textbf{3.6927} &\textbf{3.7717} &\textbf{3.8831} \\
LG-LDM+Wav2Sem$_c$   &3.7042 &3.7863 &3.9014 \\
\hline
\end{tabular}
}
\label{ablation}
\end{table}

\textbf{Comparison with BERT}
As shown in Figure \ref{fablation}, We compare the semantics learned by BERT~\cite{bert} in textual modality with our Wav2Sem learned in audio modality. Extensive experiments on the VOCASET~\cite{VOCA} for six representative speech-driven facial animation methods are conducted, and the results are reported in Table \ref{ablation}. We find that the results of integrating Wav2Sem significantly outperform those of BERT. We conduct a unified sampling of BooksCorpus and Librishpeech and visualize their distribution along with VOCASET in the BERT semantic space, as shown in Figure \ref{userstudy} (a). We notice that the data distribution of VOCASET testing is more biased towards Librisppeech, and the domain gap between these datasets leads to Wav2Sem performing better than BERT. Meanwhile, Wav2Sem extracts key knowledge and information from BERT through refinement while reducing redundant complexity. This enables Wav2Sem to be more robust and generalize effectively even with smaller data volumes, leading to better performance in downstream tasks. In addition, a significant advantage of Wav2Sem is that it does not need to rely additionally on textual modalities as input like BERT, which makes it more independent and flexible in audio processing tasks. By focusing on the audio signal itself, Wav2Sem is able to fully exploit the semantic information in the audio, improving the model's performance in various tasks. This characteristic not only simplifies the input requirements but also provides greater convenience for practical applications in audio recognition and understanding.

\subsection{User study}
To evaluate the generated effects of different audio-driven methods with and without Wav2Sem, We recruit 24 users to evaluate the results on VOCASET and BIWI datasets with a 5-point scale. Higher scores represent higher quality. Statistic results are shown in Figure \ref{userstudy} (b). The data shows that the quality of generated homophones by different methods has significantly improved by injecting Wav2Sem.

\subsection{Phoneme Recognition}
Self-supervised audio encoders are widely used in many downstream tasks. The improvements brought by our Wav2Sem are not limited to the speech-driven facial animation task. To verify the transferability of audio semantic information, we apply Wav2Sem in the phoneme recognition task to further evaluate its generality in feature decoupling.

We finetune existing self-supervised audio models, Wav2Vec 2.0~\cite{wav2vec} and HuBERT~\cite{hubert}, using labeled TIMIT data 
and evaluate the phoneme recognition error rates (PER). As shown in Table \ref{TIMIT}, it can be found that the integration of Wav2Sem can effectively enhance the accuracy of Wav2Vec 2.0 and HuBERT in phoneme recognition, which verifies the effectiveness of optimizing phoneme level features by learning semantic information in audio.
\begin{table}[htp]
\centering
\caption{Results of phoneme recognition regarding phoneme recognition error rates (PER) on TIMIT dataset.}
\setlength\tabcolsep{4pt}
\vspace{-2mm}

\begin{tabular}{ccc}
\hline
Method & dev PER & test PER \\
\hline
Wav2Vec 2.0~\cite{wav2vec} &7.4  & 8.3 \\
Wav2Vec 2.0 + Wav2Sem$_{m}$ &7.3  & 8.2 \\
Wav2Vec 2.0 + Wav2Sem$_{c}$ &\textbf{7.1}  & \textbf{8.0} \\
\hline
HuBERT~\cite{hubert}     &6.9  & 8.0 \\
HuBERT + Wav2Sem$_{m}$     &6.7  & 7.9 \\
HuBERT + Wav2Sem$_{c}$    &\textbf{6.8}  & \textbf{7.8} \\
\hline
\end{tabular}
\label{TIMIT}
\end{table}
\section{Conclusion}
In this work, we discover that existing audio self-supervised models exhibit significant coupling phenomena in the feature space when handling near-homophonic syllables, leading to an averaging effect in the generation of lip movements. To address this issue, we propose the plug-and-play Wav2Sem, which can capture the semantic information from audio signals. By leveraging global semantic information, Wav2Sem decouples the self-supervised audio encoding features in the feature space, resulting in more expressive audio representations. We have experimented with a variety of self-supervised audio models and 3D speech-driven facial animation frameworks and extensive 
experiments demonstrate the effectiveness of Wav2Sem.

\section*{Acknowledgement}
This research is supported in part by National Key R${\&}$D Program of China (No. 2022ZD0115902), the Major Key Project of PengCheng Laboratory (PCL2023A10-2), Beijing Natural Science Foundation(4232023), China Ministry of Education Funds for Humanity, Social Sciences (24YJCZH458) and the Pioneer Centre for AI, DNRF grant number P1.

{
    \small
    \bibliographystyle{ieeenat_fullname}
    \bibliography{main}
}


\end{document}